\documentstyle[preprint,aps]{revtex}

\begin{document}

\begin{center} The Effect of Deformation on the Twist Mode \medskip \\ 
Shadow J.Q. Robinson and Larry Zamick

\noindent  Department of Physics and Astronomy,\\
Rutgers University, Piscataway, \\New Jersey  08854-8019
\end{center}

\bigskip
\begin{abstract}
Using $^{12}$C as an example of a strongly deformed nucleus we 
calculate the strengths and energies in the asymptotic (oblate) 
deformed limit for the isovector twist mode operator 
$[rY^{1}\vec{l}]^{\lambda=2}t_{+}$ where l is 
the orbital angular momentum.  We also consider the $\lambda =1$ 
case.  For $\lambda=0$, the operator vanishes.  Whereas in a 
$\Delta N=0$ Nilsson model the summed strength is independent 
of the relative P$_{3/2}$ and P$_{1/2}$ occupancy when we allow 
for different frequencies $\omega_{i}$ in the x, y, and z directions 
there is a weak dependency on deformation.
\end{abstract}
\vspace{2.5in}

\newpage

\section{Introduction}

In previous works we showed that at the $\Delta N=0$ Nilsson level 
the summed strengths for the various isovector dipole modes rY$^{1}$t, 
$[rY^{1}s]^{\lambda}$t, and [rY$^{1}\vec{l}]^{\lambda}$t were 
independent of deformation, or more precisely of the valence occupancy.  
This is in striking contrast to the scissors mode excitation 
induced by the operator $ L_{\pi}-L_{\nu}$.  For the latter operator, 
the summed strength 
for $^{12}$C increased by a factor of three in going from the 
spherical limit to the $\Delta N=0$ asymptotic(oblate) limit.  Likewise, the spin
magnetic moment dipole strength shows a great sensitivity to 
deformation.  The strength is quite large in the spherical 
limit but vanishes in the asymptotic (oblate) limit.  The $\Delta N=1$ 
strength for the stretched state in $^{12}$C will obviously 
depend on the details of the P$_{3/2}$ and P$_{1/2}$ occupancy.  
You cannot excite the stretched 4$^{-}$ state in $^{12}$C by exciting 
a P$_{1/2}$ nucleon to the 1s-0d shell.  Hence the summed strength 
is proportional to the P$_{3/2}$ occupancy.  

In this work we remove the $\Delta N =0 $ restrictions and by 
using deformed harmonic oscillator wavefunctions in order to obtain the 
deformation dependence i.e. the dependence on $\delta$ of the above 
dipole modes.  We will especially focus on the 'twist mode' 
an M2 orbital mode which has been given a picturesque geometric 
description by Holzwarth et. al [1].  The problem of separating 
the M2 orbital strength from the M2 spin strength has been 
discussed by Richter [2] in the context of a detailed calculation 
by Drozdz et. al [3].  We shall consider also consider the mode 
$[rY^{1}\vec{l}]^{\lambda=1}$.  Note that the operator $[rY^{1}\vec{l}]^{\lambda=0}$ is zero.

\section{The calculation for $^{12}$C}

We are considering the asymptotic oblate limit for $^{12}$C.  
The occupied states can be described by the number of quanta 
in the x,y, and z directions $(N_{x},N_{y},N_{z})$ these are 
(0,0,0) (1,0,0) and (0,1,0).  The wave functions are \\

\begin{tabular}{cc}
(0,0,0) &$ \frac{N_{(0,0,0)}}{\sqrt{b_{x}b_{y}b_{z}}}
exp(-\frac{x^{2}}{2b_{x}^{2}}-\frac{y^{2}}{2b_{y}^{2}}-\frac{z^{2}}{2b_{z}^{2}}) $\\
(1,0,0) & $\frac{N_{(1,0,0)}}{\sqrt{b_{x}b_{y}b_{z}}}\frac{x}{b_{x}}
exp(-\frac{x^{2}}{2b_{x}^{2}}-\frac{y^{2}}{2b_{y}^{2}}-\frac{z^{2}}{2b_{z}^{2}}) $\\
\end{tabular}\\

where \\
$N_{(0,0,0)}=1/\pi^{3/2}$\\
$N_{(1,0,0)}=2/\pi^{3/2}$ etc\\
In the above $b_{x}$ is the oscillator length parameter
$b_{x}^{2}=\hbar/m\omega_{x}$ e.t.c.

Except for the fact that we now use deformed oscillator 
wave functions the calculations proceed in the same manner 
as described in the previous publications [4,5]

\section{Previous Results}

In the $\Delta N=0$ calculations we found that the summed 
strength for the electric dipole operator $rY^{1}t_{+}$ was 
$4 \pi SUM / b^{2}=27$.  
For the spin dipole the sum was 20.25 which was shown to 
be 27 $\vec{s}\cdot \vec{s}$ where of course $\vec{s} \cdot 
\vec{s}$ = 3/4 for a single nucleon.  For the orbital 
dipole $[Y1\vec{l}]^{\lambda}t_{+}$ the summed 
strength (including a sum over $\lambda$) was 48 which could 
be understood as the dipole value multiplied by $\vec{l} \cdot \vec{l}$.  
For the p shell the ordinary dipole contribution is 24 and 
$\vec{l} \cdot \vec{l}$ is two - this explains the result of 48.

A further breakdown of the orbital dipole into the $\lambda$ 
components is the second work on this subject [4,5] that gave 
the following results both in the spherical and asymptotic (oblate) limits.\\

\begin{tabular}{cc}
$\lambda$ & $4 \pi SUM$\\
0       &  0 \\
1       & 18\\
2       & 30\\
\end{tabular}
\\

For $\lambda =0 $ the operator simply vanishes.  It is proportional 
to $\vec{r} \cdot \vec{l}$ and one does not have any component of the angular 
momentum along the radial direction.  Note that the $\lambda = 2$ 
to $\lambda =1$ strength is in the ratio 
$(2\lambda+1)_{\lambda=2}/(2\lambda+1)_{\lambda=1}=30/18$.

We use the results in this section as taking off points.  In 
the next section, we introduce explicit deformation effects coming 
from the fact that $\omega_{x}$ and $\omega_{z}$ (or b$_{x}$ and b$_{z}$) are 
different.  Of course the results that we obtain should reduce 
to the ones in this section when the frequencies in the x, y, 
and z direction are taken as equal to each other.

\section{Deformed Oscillator Model with Mottelson Conditions: }

We use the Mottelson conditions to get the deformation 
parameters for $^{12}$C.  They are
\begin{equation}
\Sigma_{x}\omega_{x}=\Sigma_{y}\omega_{y}=\Sigma_{z}\omega_{z}
\end{equation}
where $\Sigma_{x}=Sum(N_{x}+1/2)$ where for a given state N$_{x}$ 
is the number of quanta in the x direction etc.
The occupied states $(N_{x},N_{y},N_{z})$ are (0,0,0) , 
(1,0,0), and (0,1,0) giving values of 
$\Sigma_{x}=10$, $\Sigma_{y}=10$, and $\Sigma_{z}=6$.  This yields 
the correct result that $^{12}C$ is oblate and strongly deformed.  We also have 
\begin{equation}
\hbar\omega_{x}=\hbar\omega_{y}=\frac{6}{10} \hbar\omega_{z}
\end{equation}
We also introduce the deformation parameter $\delta$ (which Bohr and 
Mottelson called $\delta_{OSC}$) defined by 
\begin{eqnarray}
(\hbar\omega_x)^{2}=(\hbar\omega_{0}(\delta))^{2}(1+\frac{2}{3}\delta)\\
(\hbar\omega_z)^{2}=(\hbar\omega_{0}(\delta))^{2}(1-\frac{4}{3}\delta)
\end{eqnarray}
 
The volume conservation condition is 
\begin{equation}
\hbar\omega_{x}\hbar\omega_y\hbar\omega_z=(\hbar\bar{\omega_{0}})^{3}
\end{equation}
which leads to 
\begin{equation}
\hbar\omega_{0}(\delta)=[(1+\frac{2}{3}\delta)^{2}(1-\frac{4}{3}\delta)]^{-1/6}\hbar\bar{\omega_{0}}
\end{equation}
We choose $\hbar\bar{\omega_{0}}$ (which is independent of $\delta$) to be 
15 MeV. We then get the following values
\begin{eqnarray}
\hbar\omega_{0}=15.966MeV\\
\hbar\omega_{x}=12.651MeV\\
\hbar\omega_{z}=21.086MeV
\end{eqnarray}
The oscillator length parameters have the following values
\begin{eqnarray}
b_{x}^{2}=3.2775 \\
b_{z}^{2}=1.9665\\
\bar{b_{0}}^{2}=2.7643
\end{eqnarray}

\section{Discussion of Results}

The results for the energies and excitation strengths for the
$\lambda =2$ twist mode are given in Table I, and for the
corresponding $\lambda=1$ mode in Table II.
We see that the main effect of deformation on the
$\lambda=2$ twist mode is not so much to increase
the overall strength, but rather to redistribute it.
It is true that when we go from the spherical limit
to the asymptotic deformed limit the strength, or
more precisely $4\pi strength/\bar{b_{0}}^{2}$ increases
from 30 to 38.784, a 30 percent rise but this is small compared
to the corresponding factor of three increase for the
scissors mode.  However, as shown in Table I, whereas in
the spherical oscillator limit all the strength would be
at an excitation energy of $\hbar\bar{\omega_{0}}=$ 15 MeV,
we now have the $\Delta N=1$ strength split into three
parts at 12.651, 21.086, and 29.251 MeV.  There is also
some $\Delta N =3$ strength at 46.388 MeV and 54.823 MeV.
For the $\lambda =1$ mode there is a similar redistribution
of strength, as seen in Table II.

\section{Expressions in terms of the Deformation parameter $\delta$}

For the $\lambda= 2$ (TWIST) mode up to second order 
in $\delta$, the values of $4\pi/b^{2}$ summed strengths for 
$\delta = -0.55814)$ are for $\Delta N =1$
\begin{equation}
30(1-\frac{2}{15} \delta+\frac{133}{360}\delta^{2})=35.685
\end{equation}
and for $\Delta N=3$
\begin{equation}
\frac{33}{2} \delta^{2}=5.140
\end{equation}

Note that for $\lambda=2$, $\Delta N =1$ there are terms linear in $\delta$ 
but not for $\lambda=2$,  $\Delta N =3$ 
We find that at this order the $\Delta N=1$ result is good 
(35.685 compared to exact value of 35.558) but for the 
$\Delta N=3$ result we would need to consider higher 
order terms for such a large deformation. (5.140 compared to exact value of 3.226)

For the $\lambda=1$ mode up to second order in $\delta$ we obtain
\begin{eqnarray}
18 (1 + 35/72\delta^{2})=20.725\\
\frac{21}{2} \delta^{2}=3.271
\end{eqnarray}

In contrast to the $\lambda=2$ case, for $\lambda =1$ there 
are no terms linear in $\delta$.
Again for the $\Delta N=1$ the expansion is good
(20.725 compared to exact value of 20.156) but for the
$\Delta N=3$ result we would need to consider higher
order terms for such a large deformation. (3.271 compared to exact value of 2.751)

\section{Comparison with other mode - Energy Weighted Sum Rule}

The twist excitation operator is $[rY^{1}\vec{l}]^{\lambda =2 }t$.  
The scissors mode operator is $(\vec{L_{\pi}}-\vec{L_{\nu}})$. 
The latter mode will not be excited unless there are open shells 
of both neutrons and protons.  The twist mode on the other hand 
can be excited in a closed shell nucleus.  Indeed measurements 
attempting to separate the orbital M2 from the spin 
M2 in $^{48}$Ca and $^{90}$Zr were performed by P. von 
Neumann-Cosel et. al. [6]  This 
already indicates that the twist mode is less sensitive to deformation 
than the scissors mode.  The scissors mode depends on deformation 
for its very existence.  As just mentioned whereas in $^{12}$C 
when we go from the spherical limit to the asymptotic (oblate) 
limit the summed scissors mode strength increase by a factor of 
three whereas the twist mode strength increases by only 30 percent.

We next compare the twist mode to the ordinary dipole mode $rY^{1}t$.  
For the latter mode even with deformation there will be no 
$\Delta N =3$ excitation for our simple deformed oscillator hamiltonian 
$P^{2}+\frac{1}{2}m \omega_{x} x^{2} +\frac{1}{2}m\omega_{y}y^{2}+\frac{1}{2}m\omega_{z}z^{2}$.  
This is because the dipole operator components are basically x, y, 
and z and as such can only excite one quantum.  

Whereas in the case of axial symmetry the twist mode is split into 
three $\Delta N =1$ parts and two $\Delta N =3 $ parts, the ordinary 
dipole is split only into two parts at excitation energies of $\hbar\omega_{x}$
and $\hbar\omega_{z}$, with twice the strength in the former.  

The energy weighted 
strength for the ordinary dipole does not change as we go from spherical 
to the asymptotic (oblate) limit.  This is because the dipole operator 
components x, y, and z commute with the potential energy term 
$\frac{1}{2}m\omega_{x}x^{2}+\frac{1}{2}m\omega_{y}y^{2}+\frac{1}{2}m\omega_{z}z^{2}$.  
Furthermore the ordinary strength does not change either.  
We can see this by the fact that the ordinary strength is proportional 
to $\frac{1}{\omega_{x}}+\frac{1}{\omega_{y}}+\frac{1}{\omega_{z}}$.  
When we calculate the EWSR we get rid of these energy factors.

The Energy Weighted Strengths for the twist mode 
and corresponding $\lambda=1$ mode on the other hand
are much greater in the deformed limit than in the spherical limit.  
For $\lambda=2$ the corresponding values are 854.136 MeV and 450 MeV, 
while for $\lambda=1$ they are 516.622 MeV and 270 MeV.  The reason 
for this is that the operator $[rY^{1}\vec{l}]^{\lambda}$ does 
not commute with the potential energy terms unless all three 
frequencies $\omega_{x}$, $\omega_{y}$, and  $\omega_{z}$ are the same.

This work was supported by the U.S. Dept. of Energy under Grant No. DF-FG02-95ER-40940

\begin{table}
\caption{Results for Operator $[rY^{1}\vec{l}]^{\lambda=2}$ (Twist Mode)}
\begin{tabular}{ccc}
\tableline
a) Symbolic  &   &\\
\tableline
Excitation Energy & 4$ \pi$ Strength&\\
\tableline
$\hbar\omega_{x}		    $ &$ \frac{114 b_{x}^{4}+18 b_{z}^{4}
-60b_{x}^{2}b_{z}^{2}}{8b_{x}^{2}}$&\\
$\hbar\omega_{z}      	    $ &$ 9b_{z}^{2}+6 \frac {b_{x}^{4}}{b_{z}^{2}}     $ &\\ 
$2\hbar\omega_{z}-\hbar\omega_{x}$ & $\frac{6b_{z}^{4}
+6b_{x}^{4}+12b_{z}^{2}b_{x}^{2}}{4b_{x}^{2}} $ &\\
$\Delta N =1$ Sum &$\frac{9}{2}b_{z}^{2}+\frac{63}{4}b_{x}^{2}
+\frac{15}{4}\frac{b_{z}^{4}}{b_{x}^{2}}+ 6 \frac{b_{x}^4}{b_{z}^{2}}$ & \\
$2\hbar\omega_{x}+\hbar\omega_{z}$ & $\frac{15}{2}b_{z}^{2}-15b_{x}^{2}+\frac{15}{2}\frac{b_{x}^{4}}{b\_{z}^{2}}$&\\
$2\hbar\omega_{z}+\hbar\omega_{x}$ & $\frac{9}{2}b_{x}^{2}-9b_{z}^{2}
+\frac{9}{2}\frac{b_{z}^{4}}{b_{x}^{2}}$ &\\
$\Delta N = 3$ Sum & $ -\frac{39}{2}b_{x}^{2}+3b_{z}^{2}+\frac{9}{2}\frac{b_{z}^{4}}{b_{x}^{2}}+12 \frac{b_{x}^{4}}{b_{z}^{2}}$ & \\
\end{tabular}
\begin{tabular}{ccc}
\tableline
b) Numerical (Deformed Oscillator) & &\\
\tableline
Excitation Energy(MeV) &$4  \pi Strength/\bar{b_{0}}^{2}$ 
& Spherical limit (4$\pi Stength/\bar{b_{0}}^{2}$)\\
\tableline
12.651 & 12.520 & 9\\
21.086 & 18.259 & 15\\
29.521 &  4.779 & 6\\
$\Delta N=1$ Sum & 35.558  & 30\\
46.388 & 2.372 & 0\\
54.823 & 0.854 & 0\\
$\Delta N=3$ Sum & 3.226 & 0\\
\tableline
\end{tabular}
\end{table}

\begin{table}
\caption{Results for Operator $[rY^{1}\vec{l}]^{\lambda=1}$}
\begin{tabular}{ccc}
\tableline
a) Symbolic  &   &\\
\tableline
Excitation Energy & 4 $\pi$ Strength&\\
\tableline
$\hbar\omega_{x}                 $ &$ \frac{33}{4}b_{x}^{2}
-\frac{3}{2}b_{z}^{2}+\frac{9}{4}\frac{b_{z}^{4}}{b_{x}^{2}}$&\\
$\hbar\omega_{z}                 $ &$ 3\frac{b_{x}^{4}}{b_{z}^{2}}$&\\
$2\hbar\omega_{z}-\hbar\omega_{x}$ &$ \frac{3}{2}b_{x}^{2}+3b_{z}^{2}
+\frac{3}{2}\frac{b_{z}^{4}}{b_{x}^{2}}   $&\\
$\Delta N=1$ Sum &$3 \frac{b_{x}^{4}}{b_{z}^{2}}
+\frac{39}{4}b_{x}^{2}+\frac{3}{2}b_{z}^{2}+\frac{15}{4}\frac{b_{z}^{4}}{b_{x}^{2}}$   & \\
$2\hbar\omega_{x}+\hbar\omega_{z}$ &$ 6b_{z}^{2}+6\frac{b_{x}^{4}}{b_{z}^{2}}-12b_{x}^{2}$&\\
$2\hbar\omega_{z}+\hbar\omega_{x}$ &$ \frac{9}{2}b_{x}^{2}
+\frac{9}{2}\frac{b_{z}^{4}}{b_{x}^{2}}-9b_{z}^{2}$   &\\
$\Delta N=3$ Sum &$-\frac{15}{2}b_{x}^{2}-3b_{z}^{2}+\frac{9}{2}\frac{b_{z}^{4}}{b_{x}^{2}}+6\frac{ b_{x}^{4}}{b_{z}^{2}}$  &  \\
\end{tabular}
\begin{tabular}{ccc}
\tableline
b) Numerical (Deformed Oscillator) & &\\
\tableline
Excitation Energy(MeV) &$4  \pi Strength/\bar{b_{0}}^{2}$ & Spherical limit (4$\pi$ Strength/$\bar{b_{0}}^{2}   $)\\
\tableline
12.651 &  9.675 & 9\\
21.086 &  5.928 & 3\\
29.521 &  4.553 & 6\\
$\Delta N =1$ Sum & 20.156 & 18\\
46.388 &  1.897 & 0\\
54.823 &  0.854 & 0\\
$\Delta N =3$ Sum & 2.751 & 0\\
\tableline
\end{tabular}
\end{table}

\newpage
\centerline{References}
\bigskip

\begin{enumerate}
\item G. Holzwarth and G. Eckart, Z. Phys. \underline{A283} (1977) 219; Nucl. Phys. A 325 (1979) 1

\item A Richter, Progress in Particle and Nuclear Physics \underline{44} (2000) 3

\item S. Drozdz, S.Nishizaki, J. Speth and J Wambach, Phys. Rep. \underline{197} (1990) 1

\item L. Zamick and N. Auerbach, Nuclear Physics A \underline{658}, (1999) 285

\item S.J.Q. Robinson, L Zamick, A. Mekjian, and N. Auerbach Phys. Rev \underline{C62},(2000) 017302

\item P. von Neumann-Cosel et. al. , Phys. Rev. Lett. \underline{82} (1999) 1105
\end{enumerate}

\end{document}